\documentclass[11pt,draftclsnofoot,onecolumn]{IEEEtran}

\usepackage{amssymb}
\usepackage{amsmath, amsthm}
\usepackage[dvips]{graphicx}
\usepackage[font=footnotesize]{subfig}
\usepackage{kbordermatrix}

\renewcommand{\kbldelim}{(}
\renewcommand{\kbrdelim}{)}

\begin{document}

\title{Adaptive MAC Protocols Using Memory\\ for Networks with Critical Traffic}

\author{Jaeok Park and Mihaela van der Schaar\thanks{The authors are with Electrical Engineering Department, University of
California, Los Angeles (UCLA), 420 Westwood Plaza,
Los Angeles, CA 90095-1594, USA. e-mail:
\{jaeok, mihaela\}@ee.ucla.edu.}}

\maketitle

\begin{abstract}
We consider wireless communication networks where network users are
subject to critical events such as emergencies and crises.
If a critical event occurs to a user, the user needs to send
critical traffic as early as possible. However, most existing medium
access control (MAC) protocols are not adequate to meet the urgent need
for data transmission by users with critical traffic. In this paper, we devise
a class of distributed MAC protocols that achieve coordination
using the finite-length memory of users containing their own observations and
traffic types. We formulate a protocol design problem and find optimal
protocols that solve the problem. We show that the proposed protocols enable
a user with critical traffic to transmit its critical traffic
without interruption from other users after a short delay while allowing
users to share the channel efficiently when there is no critical
traffic. Moreover, the proposed protocols require short memory and can
be implemented without explicit message passing.
\end{abstract}

\begin{IEEEkeywords}
Adaptive protocols with memory, distributed medium access control protocols, networks with critical traffic,
slotted multiaccess communication.
\end{IEEEkeywords}

\section{Introduction}

Network users may face critical events where life or livelihood
is at risk. Examples include a fire in a building, a natural
disaster in a region, a heart attack of a patient, and a military
attack by an enemy. When a network user detects a critical event, it
is important for the user to inform relevant rescue parties of the
event as early as possible so that they can take necessary
measures to mitigate the risk or help affected parties recover.
The goal of this paper is to devise a medium access control (MAC)
protocol that achieves a small delay in transmitting information
about critical events, or critical traffic, in a distributed wireless communication network.
Since critical events do not occur frequently, especially when the network size
is small, it is also important to maintain good performance in terms of
throughput and fairness
when there is no critical traffic in the network.

In a network with critical traffic, a MAC protocol needs to achieve
two kinds of coordination: (i) coordination between a user with critical
traffic and other users in the case of a critical event and (ii) coordination
among users when there is no critical traffic. These two kinds of coordination
can be easily achieved if message passing is allowed.
In the case of a critical event, the user with critical traffic
can be given priority by broadcasting its traffic type to induce other users
to wait while critical traffic is transmitted.
Also, when there is no critical traffic, users can share the communication medium
in a contention-free manner by using coordination messages from a central controller
as in time division multiple access (TDMA).
However, explicit message passing is costly and often impractical in
a distributed network environment, which makes achieving coordination a challenging task.

In this paper, we aim to achieve coordination without explicit message passing
by using an extension of MAC protocols with memory, formulated in \cite{park}.
\cite{park} considers a stationary setting without critical events
and investigates how utilizing memory can help users achieve coordination
of the second kind in a distributed environment. With a protocol with memory, users determine their transmission
probabilities based on the finite-length history of their own observations
(transmission actions and feedback information). As users take transmission actions
in a probabilistic manner, the history of users evolves differently across users as time passes.
Using the variations in the history of users as a coordination
device, we can obtain some degree of coordination without relying on explicit message passing.

The setting considered in this paper is stochastic in that the traffic types of
users are determined by an exogenous event. In order to achieve
a small delay in transmitting critical traffic (i.e., coordination of the first kind), we need to treat
users with different traffic types in a different way. Thus, we extend protocols
with memory so that transmission probabilities adjust not only
to the history of observations but also to the history of traffic types.
The proposed protocols are adaptive because users can change the modes of operation based on
their traffic types. Adaptive protocols with memory proposed in this
paper have the following properties:
\begin{enumerate}
\item (coordination in a critical phase) When a critical event occurs, the user with critical traffic
captures the channel after a small delay while other users wait
until critical traffic is completely transmitted. Furthermore,
a delay constraint can be imposed to guarantee the average delay
below a certain threshold level.
\item (coordination in a normal phase) When there is no critical traffic, a success period
and a contention period are alternated. A success period contains consecutive successes
by a single user while a contention period selects a successful user
for the following success period equally likely among all users. The average
duration of a success period can be made arbitrarily large (at the expense
of reduced short-term fairness) without affecting the average duration of a contention period.
\item (no explicit message passing) The proposed protocols can be implemented
without explicit message passing between users or between a central controller and a user.
\item (short memory) The proposed protocols utilize finite memory
of a short length (1-slot memory at minimum), thus exhibiting low complexity.
\end{enumerate}

The proposed adaptive protocols have advantages over existing MAC protocols
in dealing with critical traffic. Distributed coordination function (DCF),
which is widely deployed in the IEEE 802.11a/b/g
wireless local area network (WLAN) \cite{ieee}, does not
differentiate users, and thus it is unable to give priority
to a user with critical traffic. Slotted Aloha \cite{rob} has
the same limitation. Users can be given different priorities
depending on their access categories in enhanced distributed channel
access (EDCA), which is deployed in IEEE 802.11e \cite{ieeee}.
EDCA specifies different contention window sizes and arbitration
interframe spaces (AIFS) to different access categories, yielding a smaller
medium access delay and more bandwidth for the higher-priority
traffic categories \cite{gu}. However, EDCA is designed to support
applications requiring quality-of-service, and a user having highest-priority
data shares the channel with other users. Thus, EDCA is not
directly applicable to networks with critical traffic, where it is desirable
to allocate the entire resource to a user with critical traffic.
P-MAC \cite{qiao} also differentiates users with different traffic classes
by specifying different contention window sizes. However, P-MAC does not
use AIFS, which creates a problem when applied to a network
with critical traffic because even a user with the highest priority
has a positive probability of collision at each transmission attempt.

The rest of this paper is organized as follows. In Section II, we describe
the system model. In Section III, we provide a formal representation of
adaptive protocols, define three performance metrics, and formulate
a protocol design problem. In Section IV, we provide analytical results
on how to compute the performance metrics for a given adaptive protocol.
In Section V, we solve the protocol design problem using numerical
methods. In Section VI, we discuss how adaptive protocols can be
enhanced by utilizing longer memory. In Section VII, we provide simulation results.
In Section VIII, we conclude the paper.

\section{System Model}

We consider a communication channel shared by $N$ contending users, or transmitter-receiver pairs.
We assume that the number of users is fixed
over time and known to users.\footnote{We investigate the case of the unknown number of users
in Section V.D.} Time is divided into slots
of equal length, and users maintain synchronized time slots.
A user always has packets to transmit and can attempt to transmit one packet in each
slot.
Due to interference, only one user can transmit successfully in a slot, and
simultaneous transmission by more than one user results in a collision.
After a user makes a transmission attempt, it learns whether the transmission is
successful or not using an acknowledgement (ACK) response.
We assume that there is no error in sending and receiving ACK signals.
While a user waits, it senses the channel to learn whether the channel is accessed or not.
Given this feedback structure, the set of the observations of a user in a slot
can be defined as $Y = \{\textrm{\emph{idle}}, \textrm{\emph{busy}}, \textrm{\emph{success}}, \textrm{\emph{failure}} \}$,
as in \cite{hamed}. The observation of user $i$, denoted by $y_i$,
is \emph{idle} if no user transmits, \emph{busy} if user $i$
does not transmit but at least one other user transmits, \emph{success} if
user $i$ transmits and succeeds, and \emph{failure} if user $i$ transmits
but fails.
Users are subject to critical events such as emergencies and
crises. If a critical event occurs to a user, the user is required to send
critical traffic such as a rescue message describing the critical event.
We assume that the length of critical traffic, measured by the number of packets needed to transmit it, is determined randomly.
We say that a user's traffic is normal if its traffic is not critical.
We use critical and normal users to refer to users with critical
and normal traffic, respectively.
We denote the type of user $i$'s traffic by $z_i$ and
the set of the types by $Z$ so that
$Z = \{\textrm{\emph{normal}}, \textrm{\emph{critical}}\}$.
We assume that the observation and traffic type of a user are its
local information. That is, users do not know the observations
and the types of other users.
Lastly, we assume that critical events occur infrequently so that there is at most one
critical user at a time in the system.\footnote{We consider the possibility of having
two critical users at the same time in Section VI.C.}

\section{Protocol Description and Problem Formulation}

\subsection{Description of Adaptive Protocols}

We restrict our attention to distributed protocols with which no control or coordination
messages are exchanged between a central controller and a user or between users.
We label slots by $t = 1,2,\ldots$ and use superscript $t$ to
denote variables pertinent to slot $t$.
The history of user $i$ in slot $t$ contains all the information it has obtained
before making a transmission decision in slot $t$ and can be written as
\begin{align*}
h_i^t = (z_i^1, y_i^1, \ldots, z_i^{t-1}, y_i^{t-1}, z_i^t),
\end{align*}
for $t = 2,3,\ldots$, and $h_i^1 = z_i^1$. Let $H_t$ be the set of all
possible histories for a user in slot $t$. Then the set of all possible
histories for a user can be defined by $H \triangleq \cup_{t=1}^{\infty} H_t$.
A decision rule for a user can be formally
represented by a mapping from $H$ to $[0,1]$, prescribing a transmission probability following
each possible history. A protocol is defined to be a collection of
decision rules, one for each user.

In this paper, we restrict our attention to a simple class of protocols with the
following two properties.
First, we require that protocols be symmetric in the sense that it assigns the same decision
rule to every user. The symmetry requirement can be justified by noting that
symmetric protocols are easy to implement and that users in our model are \emph{ex ante} identical.
Moreover, it simplifies our analysis significantly.
Second, we require that protocols use only the most recent
observation and the current traffic type in a stationary way (i.e., independent of slot label $t$). This requirement is motivated
by a presumption that a protocol using short memory is easy to program
and validate. We call a protocol satisfying the above two requirements
an adaptive MAC protocol with 1-slot memory, or more simply, an adaptive protocol.
It can be represented by a mapping $f: Y \times Z \rightarrow [0,1]$, which
determines the transmission probability of user $i$ in slot $t$ as
\begin{align*}
p_i^t = f(y_i^{t-1}, z_i^t),
\end{align*}
for $t = 1,2,\ldots$, where we set $y_i^0 = idle$ as initialization.
Note that adaptive protocols can be regarded as an extension
of protocols with 1-slot memory \cite{park} in that adaptive protocols allow
transmission parameters to adjust to an exogenous state variable (the traffic type in this paper).
In other words, with an adaptive protocol a user can change its modes
of operation depending on its state.

\subsection{Performance Metrics}

\subsubsection{Delay in a Critical Phase}

We define a critical phase as a period that begins with an occurrence
of a critical event and ends with the completion of critical traffic
associated with the critical event.
Let $X$ be the random variable of the length of critical traffic, measured in slots.
A protocol determines
the average number of slots that a critical phase lasts, denoted
by $T_{crit}$. The delay in a critical phase is defined as
the average number of non-success slots during a critical phase,
\begin{align*}
D_{crit} = T_{crit} - E[X].
\end{align*}

\subsubsection{Channel Utilization in a Normal Phase}

We define a normal phase as a period without critical traffic, between
two critical phases.
The channel utilization (or throughput) of users in a normal phase is defined
as the proportion of time slots in which a successful transmission occurs
during a normal phase and can be expressed as
\begin{align*}
C_{norm} = \frac{\textrm{Expected number of successes in a normal phase}}{\textrm{Expected length of a normal phase}}.
\end{align*}

\subsubsection{Fairness in a Normal Phase}

Given an adaptive protocol $f$, in a normal phase a successful user
has another success in the next slot with probability $f(success, normal) (1-f(busy, normal))^{N-1}$.
The average number of consecutive successes by a user
starting from an initial success in a normal phase is denoted by $T_s$ and is given by
\begin{align*}
T_s = \frac{1}{1 - f(success, normal) (1-f(busy, normal))^{N-1}}.
\end{align*}
A protocol with a large value of $T_s$ can
be considered as unfair in the short term because it suppresses
the transmission opportunities of other waiting users once a success occurs.
Hence, we define the fairness level in a normal phase as the inverse of the average
number of consecutive successes in a normal phase,
\begin{align*} 
F_{norm} = \frac{1}{T_s} = 1 - f(success, normal) (1-f(busy, normal))^{N-1}.
\end{align*}

\emph{Remark.} \cite{ma} defines that a protocol is $M$-short-term fair if $T_s \leq M$.
\cite{park} captures short-term fairness by considering average delay, which is
defined as the average waiting time of a user until its next success
starting from an arbitrary point of time. Although the concept of average delay
is more comprehensive than the average number of consecutive successes (as delay can be
created by other reasons than consecutive successes), we use the latter in this paper because it is much simpler
to compute and is a good proxy for the former in the case of protocols with 1-slot memory.

\subsection{Protocol Design Problem}

Suppose that the protocol designer has preferences on the
three performance metrics, $D_{crit}$, $C_{norm}$, and $F_{norm}$,
and that his preferences are represented by a utility function $U$.
Noting that a protocol determines the three performance metrics, we can express
the protocol design problem as
\begin{align*}
\max_{f \in \mathcal{F}} U(D_{crit}, C_{norm}, F_{norm}),
\end{align*}
where $\mathcal{F}$ denotes the set of all adaptive protocols.
It is reasonable to assume that the protocol designer prefers
protocols that yield a small delay in a critical phase as well as a high
channel utilization and a high fairness level in a normal phase.
Thus, we assume that $U$ is decreasing in the first argument and increasing in
the second and third arguments. Since the solution to the
protocol design problem depends highly on the specification of the utility function,
we impose the following structure on the protocol and the utility function in order to
reduce the problem into a simpler one that is readily solvable.

First, in order to achieve a small value of $D_{crit}$, we focus on
adaptive protocols satisfying $f(y, critical) = 1$ for all $y \in Y$ and
$f(busy, normal) = 0$. We call such a protocol non-intrusive. A non-intrusive
protocol guarantees that once a critical user has a successful transmission, its transmission is not interrupted
by other users (with normal traffic) until it completes the transmission of its critical traffic.
Second, we assume that the protocol designer has the most preferred fairness level
in a normal phase, denoted by $\theta \in (0,1]$, regardless of the other two
performance metrics. Then setting $F_{norm} = \theta$ together with
$f(busy, normal) = 0$ implies $f(success, normal) = 1 - \theta$.
Now it remains to specify $f(idle, normal)$ and $f(failure, normal)$, which
we denote by $q$ and $r$, respectively, for notational simplicity.
The class of adaptive protocols we consider can be written as
\begin{eqnarray*} 
&f(y, critical) = 1 \textrm{ for all } y \in Y,& \nonumber \\
&f(idle, normal) = q, \ f(busy, normal) = 0,&\\
&f(success, normal) = 1 - \theta, \ f(failure, normal)=r,& \nonumber
\end{eqnarray*}
and we call a protocol in this class a $\theta$-fair non-intrusive adaptive protocol.
Lastly, we assume that the protocol designer has a threshold level $\eta > 0$
for the delay in a critical phase such that his preferences on $D_{crit}$ is
expressed entirely with a delay constraint of the form $D_{crit} \leq \eta$.
That is, only protocols with $D_{crit}$ no larger than $\eta$ are acceptable
to the protocol designer while he does not care about the exact value of $D_{crit}$
as long as the delay constraint is satisfied. Under the three restrictions,
the protocol design problem is reduced to finding a $\theta$-fair non-intrusive
adaptive protocol that solves
\begin{align} \label{eq:pdprob}
\max_{(q,r) \in [0,1]^2} C_{norm} \textrm{ subject to } D_{crit} \leq \eta.
\end{align}
In Section IV we explain how to compute $C_{norm}$ and $D_{crit}$ analytically given
a $\theta$-fair non-intrusive adaptive protocol, whereas in Section V we investigate
the solution to \eqref{eq:pdprob} using numerical illustrations.

\emph{Remark.} The operation of a $\theta$-fair non-intrusive adaptive protocol
in a normal phase is analogous to $p$-persistent CSMA \cite{bert}. Users wait
when the channel is sensed busy and transmit with probability $q$ and $r$ following
an idle slot and a collision, respectively. Since we consider saturated arrivals
where each user always has packets to transmit, we introduce $\theta$ as a stopping
probability in order to prevent a single user from using the channel exclusively.

\section{Analytical Results}

\subsection{Derivation of the Channel Utilization in a Normal Phase}

Consider a slot $t$ in which a normal phase begins. Since slot $t-1$ is the last slot
of a critical phase, there exists
a user $i$ that completed the transmission of its critical traffic in slot $t-1$. Since user $i$
had a successful transmission in slot $t-1$, we have $y_i^{t-1}=success$ and
$y_j^{t-1} = busy$ for all $j \neq i$, and thus in slot $t$ user $i$ transmits with probability
$1-\theta$ while other users wait. Hence, a normal phase begins with a success
by the user that had critical traffic in the previous critical phase with probability $1-\theta$
and with an idle slot with probability $\theta$.\footnote{If we extend adaptive protocols as
$p_i^t = f(z_i^{t-1},y_i^{t-1},z_i^t)$, we can set $p_i^t = 0$ if $z_i^{t-1} = critical$
and $z_i^t = normal$. Then all users including the user that had critical traffic wait
in the first slot of a normal phase, which makes users contend with an equal
transmission probability in the second slot.}

To compute the channel utilization in a normal phase $C_{norm}$,
we construct a Markov chain whose state space is $\{0,1,\ldots,N\}$, where state $k$ represents
transmission outcomes in which exactly $k$ users transmit.
The transition probability from state $k$ to state $k'$ in a normal phase,
$P_{\textrm{\emph{norm}}}(k'|k)$, under a $\theta$-fair non-intrusive adaptive protocol is given by
\begin{eqnarray} \label{eq:off1}
P_{\textrm{\emph{norm}}}(k'|0) &=& \binom{N}{k'} q^{k'} (1-q)^{N-k'} \quad \textrm{for $k'=0,\ldots,N$},\\
P_{\textrm{\emph{norm}}}(k'|1) &=& \left\{ \begin{array}{ll}
\theta & \textrm{for $k'=0$}\\
1-\theta & \textrm{for $k'=1$}\\
0 & \textrm{for $k'=2,\ldots,N$},
\end{array} \right. \nonumber\\ 
P_{\textrm{\emph{norm}}}(k'|k) &=& \left\{ \begin{array}{ll}
\binom{k}{k'} r^{k'} (1-r)^{k-k'}  & \textrm{for $k'=0,\ldots,k$}\\
0 & \textrm{for $k'=k+1,\ldots,N$}
\end{array} \right. ,\textrm{for $k=2,\ldots,N$}. \label{eq:off3}
\end{eqnarray}
The transition matrix of the Markov chain can be written in the form of
\begin{align*}
\mathbf{P}_{\textrm{\emph{norm}}} = \kbordermatrix{&0&2& \cdots &N-1&N& & 1\\
0& * & * & \cdots & * & * & \vrule & *\\
2& * & * & \cdots & 0 & 0 & \vrule & *\\
\vdots & \vdots & \vdots & \ddots & \vdots & \vdots & \vrule & \vdots & \\
N-1& * & * & \cdots & * & 0 & \vrule & *\\
N& * & * & \cdots & * & * & \vrule & *\\
\cline{2-8}
1& \theta & 0 & \cdots & 0 & 0 & \vrule & 1-\theta
},
\end{align*}
where the entries marked with an asterisk can be found in \eqref{eq:off1} and \eqref{eq:off3}.

The average number of consecutive successes in a normal phase, $T_s$, is determined
by the fairness level $\theta$, where the relationship is given by $T_s = 1/\theta$.
A series of consecutive successes by a user ends with an idle slot, when the successful user waits.
Let $T_{c}$ be the average number of non-success slots until the first success
starting from an idle slot during a normal phase. A normal phase can be considered as
the alternation of a success period and a contention period, which is continued until a
critical event occurs. A success period is
characterized by consecutive successes by a user, whereas a contention period
begins with an idle slot and lasts until a user succeeds. Since all users transmit
with the same transmission probability following an idle slot, each user has an equal
chance of becoming a successful user for the following success period
at the point when a contention period starts.
In other words, a contention period selects the next successful user in a nondiscriminatory way.

Let $\mathbf{Q}_{\textrm{\emph{norm}}}$ be the $N$-by-$N$ matrix in the upper-left
corner of $\mathbf{P}_{\textrm{\emph{norm}}}$. Suppose that $0 < q,r < 1$ so that all the
entries of $\mathbf{P}_{\textrm{\emph{norm}}}$ marked with an asterisk are nonzero.
Then $(\mathbf{I} - \mathbf{Q}_{\textrm{\emph{norm}}})^{-1}$ exists and is called the
fundamental matrix for $\mathbf{P}_{\textrm{\emph{norm}}}$,
when state 1 is absorbing (i.e., $\theta = 0$) \cite{grinstead}.
The average number of slots in state $k \neq 1$ starting from state 0 (an idle slot) is given by
the $(1,k)$-entry of $(\mathbf{I} - \mathbf{Q}_{\textrm{\emph{norm}}})^{-1}$. Hence,
the average number of slots to hit state 1 (a success slot) for the first time
starting from an idle slot is given by
the first entry of $(\mathbf{I} - \mathbf{Q}_{\textrm{\emph{norm}}})^{-1} \mathbf{e}$,
where $\mathbf{e}$ is a column vector of length $N$ all of whose entries are 1.
Hence, we obtain $T_{c} = [(\mathbf{I} - \mathbf{Q}_{\textrm{\emph{norm}}})^{-1} \mathbf{e}]_1$.
Note that $T_{c}$ is independent of $\theta$. That is, the average duration of a contention
period is not affected by the average duration of a success period.
The channel utilization of users in a normal phase can be computed by
\begin{align} \label{eq:psuc}
C_{norm} = \frac{T_s}{T_{c} + T_s}
= \frac{1}{\theta [(\mathbf{I} - \mathbf{Q}_{\textrm{\emph{norm}}})^{-1} \mathbf{e}]_1 + 1},
\end{align}
for $(q,r) \in (0,1)^2$.

An alternative method to compute the channel utilization in a normal phase
is to use a stationary distribution. Since $\theta \in (0,1]$, all states
communicate with each other under the transition matrix $\mathbf{P}_{\textrm{\emph{norm}}}$
for all $(q,r) \in (0,1)^2$.
Hence, the Markov chain is irreducible, and there exists
a unique stationary distribution $\mathbf{w}_{\textrm{\emph{norm}}}$,
which satisfies
\begin{align} \label{eq:wnorm}
\mathbf{w}_{\textrm{\emph{norm}}} = \mathbf{w}_{\textrm{\emph{norm}}} \mathbf{P}_{\textrm{\emph{norm}}}& \textrm{ and }
\mathbf{w}_{\textrm{\emph{norm}}} \mathbf{e} = 1.
\end{align}
Let
${w}_{\textrm{\emph{norm}}}(k)$ be the entry of $\mathbf{w}_{\textrm{\emph{norm}}}$
corresponding to state $k$, for $k = 0,1,\ldots,N$.
Then ${w}_{\textrm{\emph{norm}}}(k)$ gives the probability of state $k$ during a normal phase.
In particular, the channel utilization in a normal phase
is given by ${w}_{\textrm{\emph{norm}}}(1)$.
Since success and contention periods alternate from the beginning
of a normal phase, the stationary distribution yields the probabilities of states
for any duration of a normal phase (assuming that a normal phase lasts sufficiently longer than
$T_s+T_c$), not just the limiting probabilities as a normal phase
lasts infinitely long. By manipulating \eqref{eq:wnorm}, we can derive that
${w}_{\textrm{\emph{norm}}}(1) = C_{norm}$, whose expression is given in \eqref{eq:psuc}.

\subsection{Derivation of the Delay in a Critical Phase}

Consider a slot $t$ in which a critical phase begins.
Since a critical user always transmits under a
non-intrusive protocol, we consider a Markov chain
whose state space is $\{0,1,\ldots,N-1\}$, where state $k$ represents
transmission outcomes in which exactly $k$ normal users transmit.
The transition probability from state $k$ to state $k'$ in a critical phase,
$P_{\textrm{\emph{crit}}}(k'|k)$, under a $\theta$-fair non-intrusive adaptive protocol is given by
\begin{eqnarray} \label{eq:on}
P_{\textrm{\emph{crit}}}(k'|k) &=& \left\{ \begin{array}{ll}
\binom{k}{k'} r^{k'} (1-r)^{k-k'} & \textrm{for $k'=0,\ldots,k$}\\
0 & \textrm{for $k'=k+1,\ldots,N-1$}
\end{array} \right. ,\textrm{for $k=0,\ldots,N-1$}.
\end{eqnarray}
The transition matrix of the Markov chain can be written in the form of
\begin{align*}
\mathbf{P}_{\textrm{\emph{crit}}} = \kbordermatrix{&1&2& \cdots &N-2&N-1& & 0\\
1& * & 0 & \cdots & 0 & 0 & \vrule & *\\
2& * & * & \cdots & 0 & 0 & \vrule & *\\
\vdots & \vdots & \vdots & \ddots & \vdots & \vdots & \vrule & \vdots & \\
N-2& * & * & \cdots & * & 0 & \vrule & *\\
N-1& * & * & \cdots & * & * & \vrule & *\\
\cline{2-8}
0& 0 & 0 & \cdots & 0 & 0 & \vrule & 1
},
\end{align*}
where the entries marked with an asterisk can be found in \eqref{eq:on}.
Note that state 0, which corresponds to a success by the critical user,
is absorbing because once the critical user has a successful transmission,
its transmissions in the following slots are not interrupted by other users with normal traffic.
Hence, the delay in a critical phase under a $\theta$-fair non-intrusive adaptive protocol
is independent of the length of critical traffic
and is measured by the average number of collisions that the critical user
experiences before obtaining a successful transmission.
Let $\mathbf{Q}_{\textrm{\emph{crit}}}$ be the $(N-1)$-by-$(N-1)$ matrix in the upper-left corner of $\mathbf{P}_{\textrm{\emph{crit}}}$.
For $r \in [0,1)$, the matrix $\mathbf{I} - \mathbf{Q}_{\textrm{\emph{crit}}}$ is invertible, and
the average number of slots until the first success starting from
state $k$ is given by the $k$-th entry of
$(\mathbf{I} - \mathbf{Q}_{\textrm{\emph{crit}}})^{-1} \mathbf{e}$, for $k = 1, \ldots, N-1$.

The number of collisions that a critical user experiences in
a critical phase depends on the transmission outcome in slot $t-1$, the last
slot of the preceding normal phase.
We represent the transmission outcome of slot $t-1$ by a pair $(l,a)$, where $l$ is the number of
transmissions by users other than the user that becomes a critical user in the following
critical phase and $a$ is the transmission action of the user. We write $a = T$ if the user transmits and $a = W$ if it waits.
Suppose that the transmission outcome of slot $t-1$ is represented by $(l,a)$, for $l = 2, \ldots, N-1$.
Then the Markov chain starts from state $l$ in slot $t-1$, regardless of $a$.
Since the critical phase starts in slot $t$, the number of collisions in the critical phase does not include
the collision in slot $t-1$. Hence, the average number of collisions
until the first success in a critical phase when the preceding normal phase
ended with $l$ transmissions by users other than the critical user is given by
\begin{align*}
d(l,a) = [ (\mathbf{I} - \mathbf{Q}_{\textrm{\emph{crit}}})^{-1} \mathbf{e} ]_l - 1,
\end{align*}
for $l = 2, \ldots, N-1$ and $a = T, W$.

Suppose that the transmission outcome of slot $t-1$ is represented by $(1,a)$.
If $a = T$, then the Markov chain starts from state $1$ in slot $t-1$,
and the average number of collisions
until the first success in a critical phase when the preceding normal phase
ended with two transmissions including one by the critical user is given by
\begin{align*}
d(1,T) = [ (\mathbf{I} - \mathbf{Q}_{\textrm{\emph{crit}}})^{-1} \mathbf{e} ]_1 - 1.
\end{align*}
If $a = W$, then there is a successful user, different from the critical user
in the following critical phase, in slot $t-1$.
The successful user transmits with probability $1-\theta$ while all the other normal users wait in slot $t$.
Thus, with probability $\theta$, the critical user succeeds in slot $t$,
and with probability $1-\theta$, state 1 occurs in slot $t$, from which
it takes $[ (\mathbf{I} - \mathbf{Q}_{\textrm{\emph{crit}}})^{-1} \mathbf{e} ]_1$ collisions
on average to reach a success by the critical user. Therefore,
the average number of collisions
until the first success in a critical phase when the preceding normal phase
ended with a success by a user other than the critical user is given by
\begin{align*}
d(1,W) = \theta \cdot 0 + (1-\theta)  [ (\mathbf{I} - \mathbf{Q}_{\textrm{\emph{crit}}})^{-1} \mathbf{e} ]_1
= (1-\theta)  [ (\mathbf{I} - \mathbf{Q}_{\textrm{\emph{crit}}})^{-1} \mathbf{e} ]_1.
\end{align*}

Suppose that the transmission outcome of slot $t-1$ is represented by $(0,a)$.
If $a = T$, then the critical user in the following critical
phase has a success in slot $t-1$.
Since $f(busy, normal) = 0$, all normal users wait in slot $t$.
Thus, the critical user has another success in slot $t$, which leads to zero delay
in a critical phase when the preceding normal phase ended with a success by
the critical user, i.e.,
\begin{align*}
d(0,T) = 0.
\end{align*}
If $a = W$, then all $(N-1)$ normal users transmit with probability $q$
in slot $t$. Then
with probability $\binom{N-1}{k} q^k (1-q)^{N-1-k}$, slot $t$ contains transmission by $k$
normal users, for $k = 0, \ldots, N-1$.
With probability $(1-q)^{N-1}$ the critical user experiences no collision while
with probability $\binom{N-1}{k} q^k (1-q)^{N-1-k}$ the critical phase begins with state $k$,
for $k = 1,\ldots, N-1$. Therefore,
the average number of collisions
until the first success in a critical phase when the preceding normal phase
ended with an idle slot is given by
\begin{align*}
d(0,W) &= (1-q)^{N-1} \cdot 0 + \sum_{k=1}^{N-1} \binom{N-1}{k} q^k (1-q)^{N-1-k} [ (\mathbf{I} - \mathbf{Q}_{\textrm{\emph{crit}}})^{-1} \mathbf{e} ]_k\\
&= \sum_{k=1}^{N-1} \binom{N-1}{k} q^k (1-q)^{N-1-k} [ (\mathbf{I} - \mathbf{Q}_{\textrm{\emph{crit}}})^{-1} \mathbf{e} ]_k.
\end{align*}

As discussed in Section IV.A, the probability that the last slot of a normal phase
has $k$ transmissions is given by ${w}_{\textrm{\emph{norm}}}(k)$, for $k = 0,1,\ldots,N$.
Since we consider symmetric protocols, the probability that a particular user
is one of $k$ transmitting users is given by $k/N$. Thus, the probability
that the transmission outcome of the last slot of a normal phase is represented by $(l,a)$, denoted by $v(l,a)$, is given by
\begin{align*}
v(l,T) = \frac{l+1}{N} {w}_{\textrm{\emph{norm}}}(l+1) \textrm{ and }
v(l,W) = \frac{N-l}{N} {w}_{\textrm{\emph{norm}}}(l),
\end{align*}
for $l = 0, \ldots, N-1$. Then the delay in a critical phase
can be computed as
\begin{align*}
D_{crit} = \sum_{l \in \{0,\ldots,N-1\}, a \in \{T,W\}} v(l,a) d(l,a).
\end{align*}

\begin{figure}
\begin{center}
\includegraphics[width=0.8\textwidth]{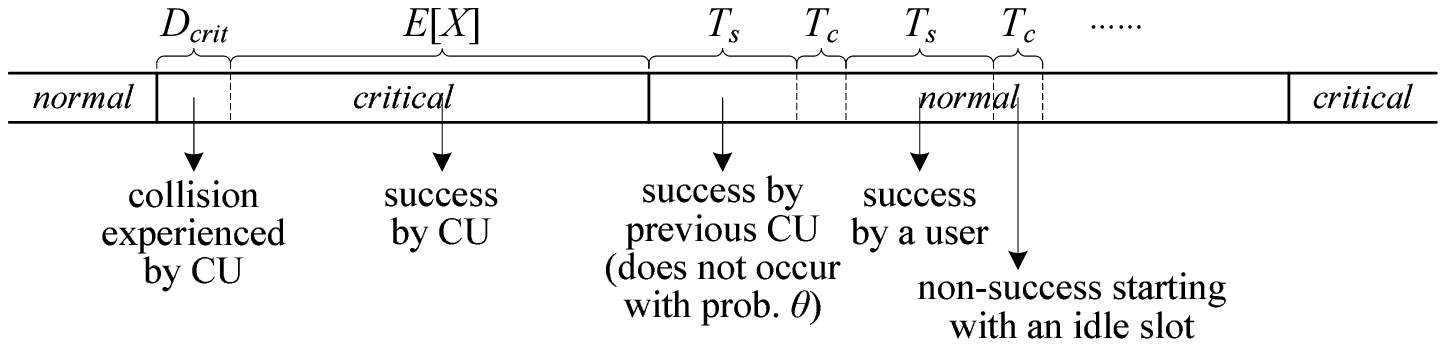}
\caption{Operation of the system under a $\theta$-fair non-intrusive adaptive protocol. (CU represents a critical user.)}
\label{fig:oper}
\end{center}
\end{figure}

Fig.~\ref{fig:oper} summarizes the operation of the system
under a $\theta$-fair non-intrusive adaptive protocol. Note that $E[X]$ is exogenously given
while $D_{crit}$, $T_{s}$, and $T_{c}$ are determined
by the protocol specification. Since $T_s$ is
determined completely by the fairness level,
the protocol design problem \eqref{eq:pdprob} can be restated as to minimize
$T_{c}$ while keeping $D_{crit}$ below a certain threshold level. Finally,
we note that we can achieve $D_{crit} = 0$ and $T_c = 0$ (and thus $C_{norm} = 1$)
for any fairness level when we allow coordination messages.
Hence, the gap between $D_{crit}$ and 0 and between $C_{norm}$ and 1 that arises
when we are restricted to use distributed protocols without message passing
can be considered as an efficiency loss due to the lack of centralized control.

\section{Numerical Results}

\subsection{Graphical Illustration of the Protocol Design Problem}

Based on the results in Section IV, we can show that, for a given fairness level $\theta \in (0,1]$,
$C_{norm}$ and $D_{crit}$ are continuous functions of $(q,r)$ on the
interior of $[0,1]^2$. In order to guarantee the existence of a solution,
in this section we consider the protocol design problem on a restricted domain,
\begin{align} \label{eq:pdprob2}
\max_{(q,r) \in [\epsilon,1-\epsilon]^2} C_{norm} \textrm{ subject to } D_{crit} \leq \eta,
\end{align}
for a small $\epsilon > 0$. Throughout this section, we set $\epsilon = 0.01$.
We say that a protocol is optimal if it solves \eqref{eq:pdprob2}.
An optimal protocol gives an approximate, if not exact, solution to \eqref{eq:pdprob}.

\begin{figure}%
\centering
\subfloat[$C_{norm}$]{%
\label{fig:cont-a}%
\includegraphics[width=0.45\textwidth]{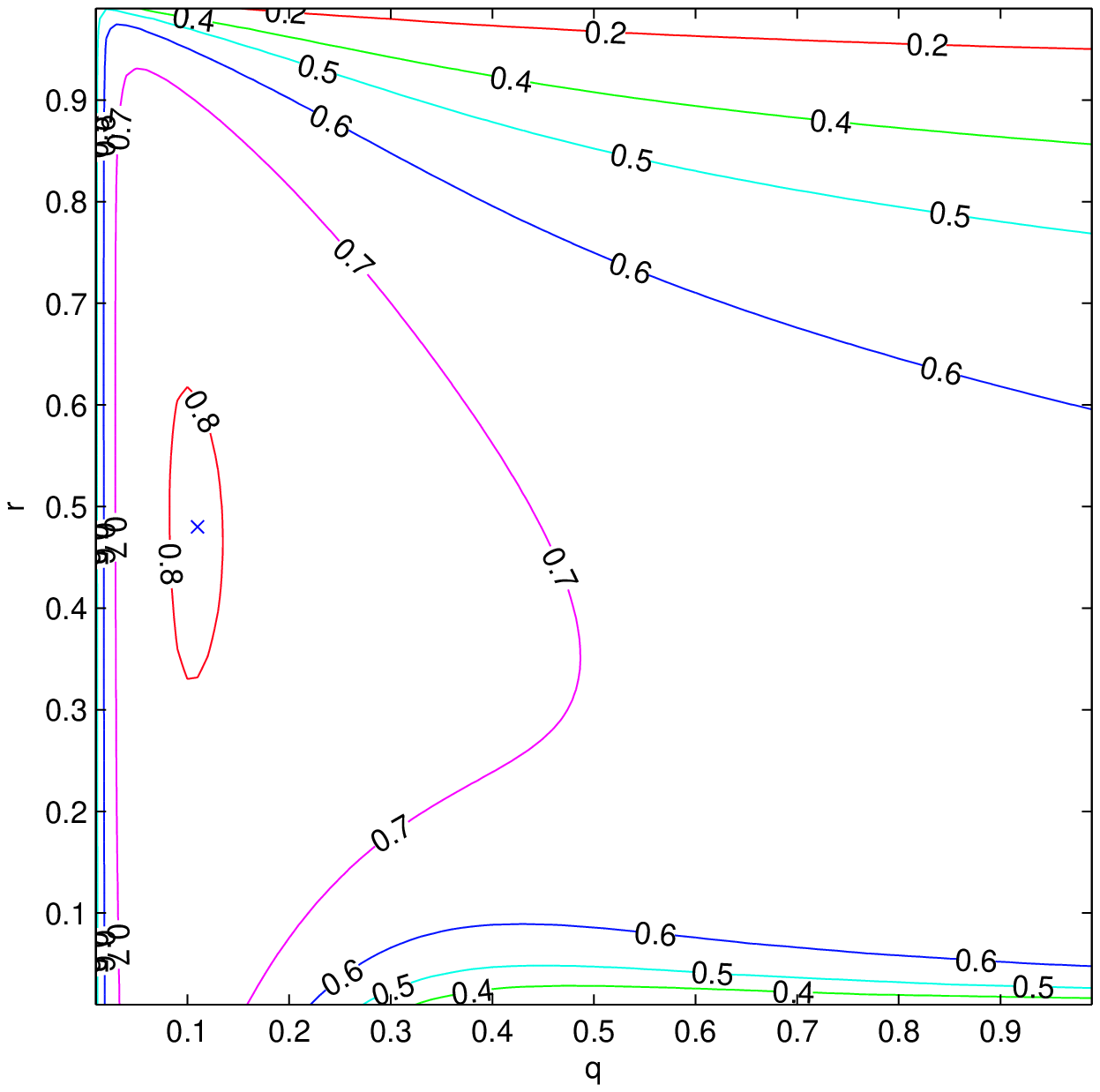}}%
\subfloat[$D_{crit}$]{%
\label{fig:cont-b}%
\includegraphics[width=0.45\textwidth]{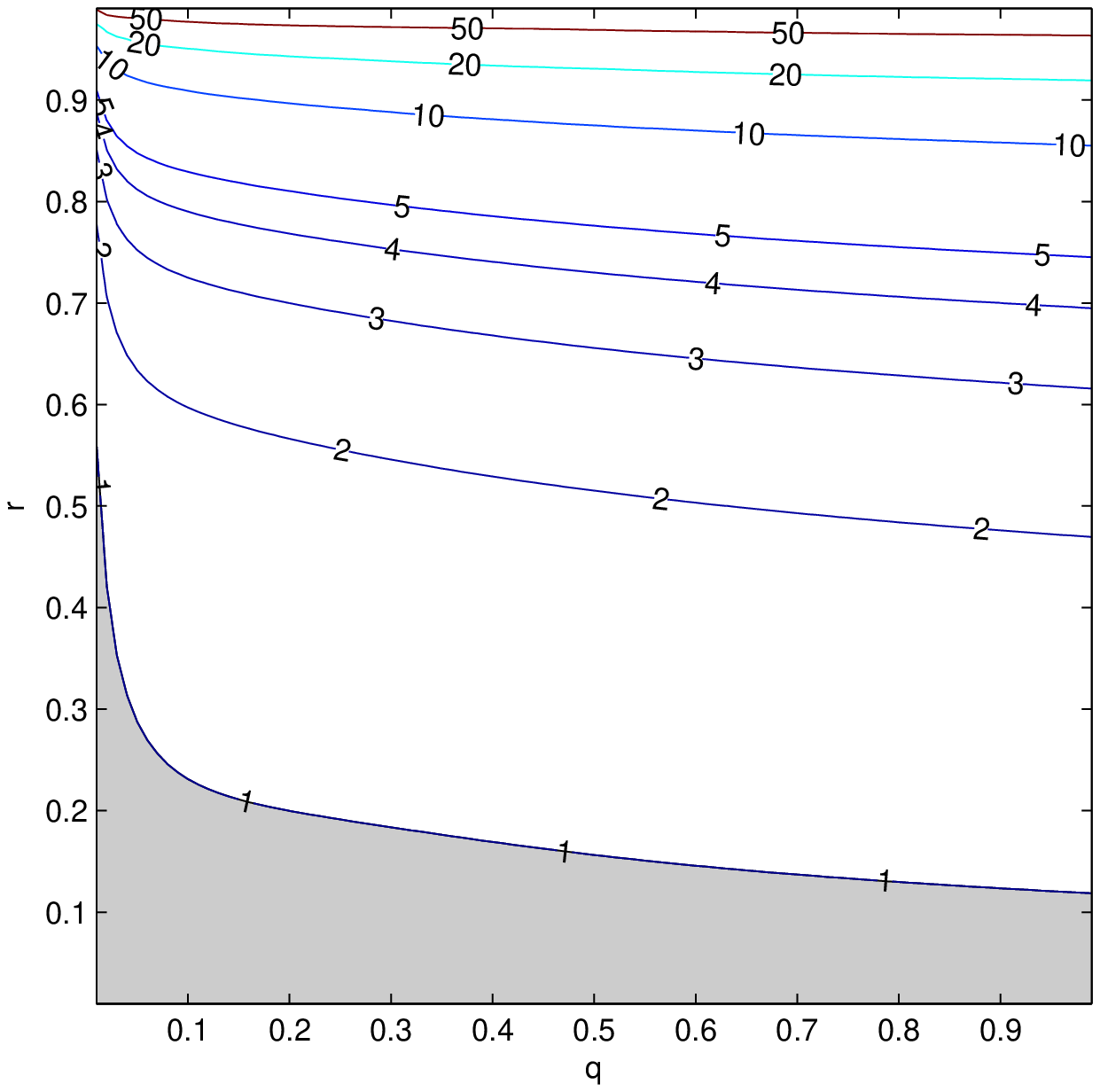}}%
\caption{Contour curves of $C_{norm}$ and $D_{crit}$ as functions of $(q,r)$
when $N = 10$ and $\theta = 0.1$.}
\label{fig:contour}%
\end{figure}

In Fig.~\ref{fig:contour}, we show the dependence of the performance metrics,
$C_{norm}$ and $D_{crit}$, on $(q,r)$. To obtain the results, we consider ten users, i.e., $N = 10$,
and fix $\theta = 0.1$ so that $T_s = 10$.
Fig.~\ref{fig:contour}\subref{fig:cont-a} plots the contour curves of $C_{norm}$.
Let $(q^*, r^*) = \arg \max_{(q,r) \in [\epsilon,1-\epsilon]^2} C_{norm}$. That is, $(q^*, r^*)$ represents
the $\theta$-fair non-intrusive adaptive protocol that maximizes the channel utilization in a normal phase when no constraint
is imposed on the delay in a critical phase. With numerical methods, we find that $(q^*, r^*)$ is unique
with the value $(0.105, 0.479)$ and achieves $0.804$ as the maximum value of $C_{norm}$.
By \eqref{eq:psuc}, $C_{norm}$ and $T_{c}$ are negatively related for a given fairness
level $\theta$, and the minimum value of $T_{c}$ corresponding to the maximum value
of $C_{norm}$ is given by $2.44$.
That is, at $(q^*, r^*)$, a contention period in a normal phase lasts for $2.44$ slots
on average, while
the average duration of a success period is given by $1/\theta$.
The value of $(q^*, r^*)$ can be justified as follows. Following an idle slot
in a normal phase, every user transmits with probability $q$, and thus the probability of success
is maximized when $q = 1/N$. During a normal phase, a collision cannot follow a success, and
following an idle slot, a collision involving two transmissions is most likely among all kinds of collisions
when $q \approx 1/N$. Since non-colliding users do not transmit following a collision under a non-intrusive protocol,
the probability of success between two contending users is maximized when $r = 1/2$.
$r^*$ is chosen slightly smaller than $1/2$ because collisions involving
more than two transmissions occur with small probability.
Fig.~\ref{fig:contour}\subref{fig:cont-b} plots
the contour curves of $D_{crit}$. As $q$ and $r$ are large,
users transmit aggressively in a contention period of a normal phase, intensifying
interference to a critical user before its first success. Thus, $D_{crit}$
is increasing in both $q$ and $r$. The set of $(q,r)$ that satisfies
a delay constraint $D_{crit} \leq \eta$ can be represented by the region below the
contour curve of $D_{crit}$ at level $\eta$. For example, the shaded area
in Fig.~\ref{fig:contour}\subref{fig:cont-b} represents the constraint set corresponding to $D_{crit} \leq 1$.

\begin{figure}
\begin{center}
\includegraphics[width=0.7\textwidth]{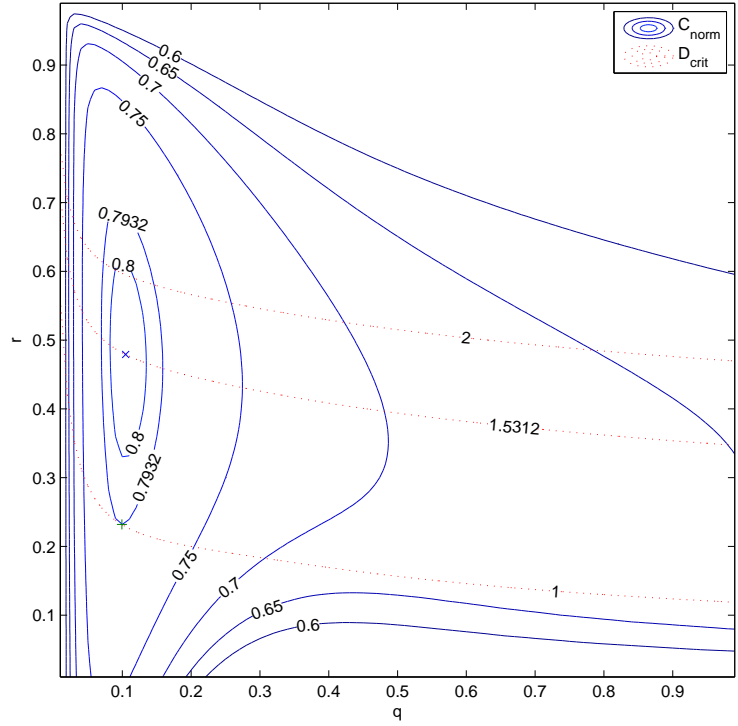}
\caption{Illustration of optimal protocols.}
\label{fig:contmap}
\end{center}
\end{figure}

Fig.~\ref{fig:contmap} shows the contour curves of $C_{norm}$ and $D_{crit}$ in
the same graph to illustrate the protocol design problem \eqref{eq:pdprob2}.
The protocol design problem is to find
the largest value of $C_{norm}$ on the region of $(q,r)$ that satisfies $D_{crit} \leq \eta$.
Let $\eta^*$ be the value of $D_{crit}$ at $(q^*, r^*)$. When $N = 10$ and $\theta = 0.1$, we have $\eta^* = 1.531$.
The constraint $D_{crit} \leq \eta$ is slack
if $\eta > \eta^*$ and is binding otherwise.
For example, if $\eta = 1$, the constraint is binding and the optimal protocol
is given by the point on the contour curve of $D_{crit}$ at level $1$, marked
with `$+$' in Fig.~\ref{fig:contmap}, where
a contour curve of $D_{crit}$ and that of $C_{norm}$ are tangent to each other.
In contrast, if $\eta = 2$, the constraint is slack and
the optimal protocol is given by the solution to the unconstrained problem, $(q^*,r^*) = (0.105, 0.479)$, marked
with `$\times$' in Fig.~\ref{fig:contmap}.

\begin{figure}%
\centering
\subfloat[][]{%
\label{fig:gvary-a}%
\includegraphics[width=0.5\textwidth]{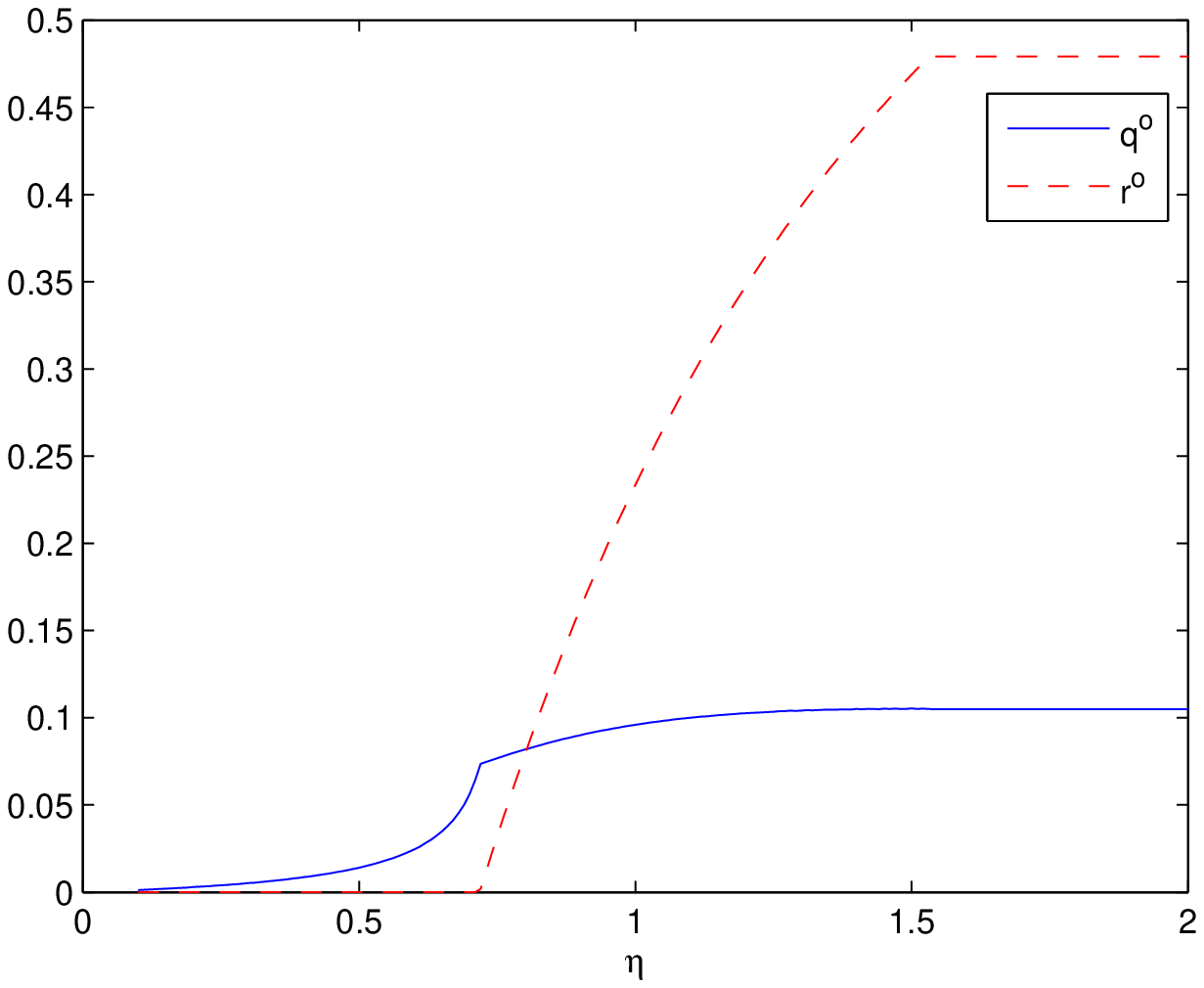}}%
\subfloat[][]{%
\label{fig:gvary-b}%
\includegraphics[width=0.5\textwidth]{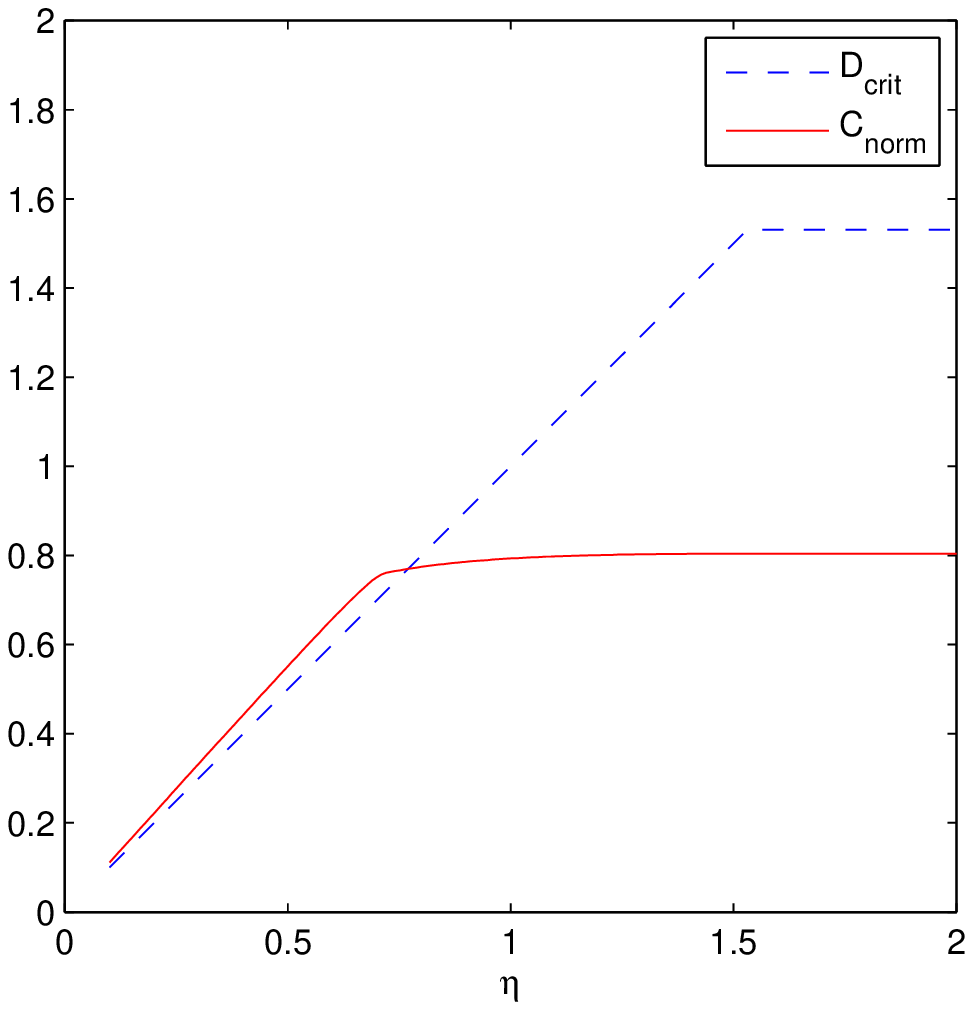}}%
\caption{Solution to the protocol design problem
for $\eta$ between $0.1$ and $2$ when $N = 10$ and $\theta = 0.1$:
\protect\subref{fig:gvary-a} optimal protocols, and
\protect\subref{fig:gvary-b} the values of $D_{crit}$ and $C_{norm}$ at the optimal protocols.}
\label{fig:gvary}%
\end{figure}

Fig.~\ref{fig:gvary} shows the solutions to the protocol design problem
for $\eta$ between $0.1$ and $2$. Fig.~\ref{fig:gvary}\subref{fig:gvary-a}
plots optimal protocols, denoted by $(q^o,r^o)$, as $\eta$ varies while
Fig.~\ref{fig:gvary}\subref{fig:gvary-b} shows the values of $D_{crit}$ and $C_{norm}$
at the optimal protocols. We can divide the range of $\eta$ into three regions:
$(0, 0.71]$, $(0.71, 1.53]$, and $(1.53, \infty)$.
For $\eta \leq 0.71$, the optimal protocol occurs at the corner with $r^o = \epsilon$. As $\eta$ decreases
in this region, $q^o$ decreases to $\epsilon$ while $r^o$ stays at $\epsilon$, which makes $C_{norm}$ decrease to 0.
Smaller $\eta$ means that higher priority is given to a critical user, and this can be achieved
by inhibiting transmissions by users when they have normal traffic. For $\eta \in (0.71, 1.53]$, the solution to
the protocol design problem is interior while the constraint $D_{crit} \leq \eta$ is still
binding. The trade-off between $D_{crit}$ and $C_{norm}$ is less severe in this region
than in $(0, 0.71]$. Reducing $\eta$ from $1.53$ to $0.71$ results in a slight decrease in $C_{norm}$
from $0.80$ to $0.76$. For $\eta > 1.53$, the constraint
$D_{crit} \leq \eta$ is slack, and thus $(q^o,r^o)$ remains at $(q^*,r^*) = (0.105, 0.479)$
while $C_{norm}$ remains at its unconstrained maximum level, $0.804$.
The rate of change in the maximum value of $C_{norm}$ with respect to $\eta$ suggests that
keeping $D_{crit}$ below $0.71$ induces a large cost in terms of the reduced channel utilization
in a normal phase, maintaining $D_{crit}$ between $0.71$ and $1.53$ only a minor cost,
and tolerating $D_{crit}$ larger than $1.53$ no cost.
In other words, when the optimal solution to the protocol design
problem is interior, the optimal dual variable on the constraint $D_{crit} \leq \eta$ is close
to zero or is zero.

\subsection{Varying the Number of Users}

\begin{figure}%
\centering
\subfloat[][]{%
\label{fig:nunc-a}%
\includegraphics[width=0.5\textwidth]{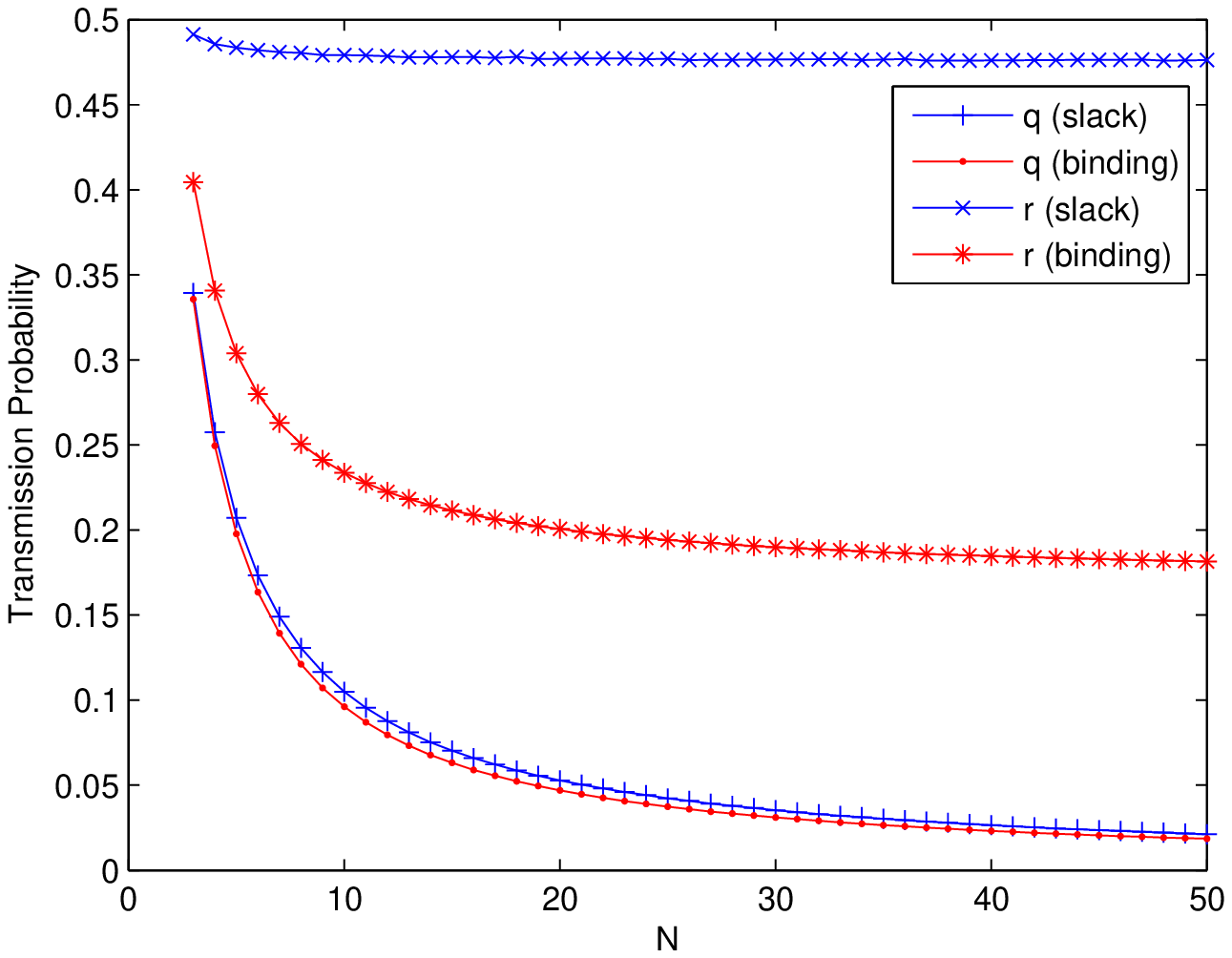}}%
\subfloat[][]{%
\label{fig:nunc-b}%
\includegraphics[width=0.5\textwidth]{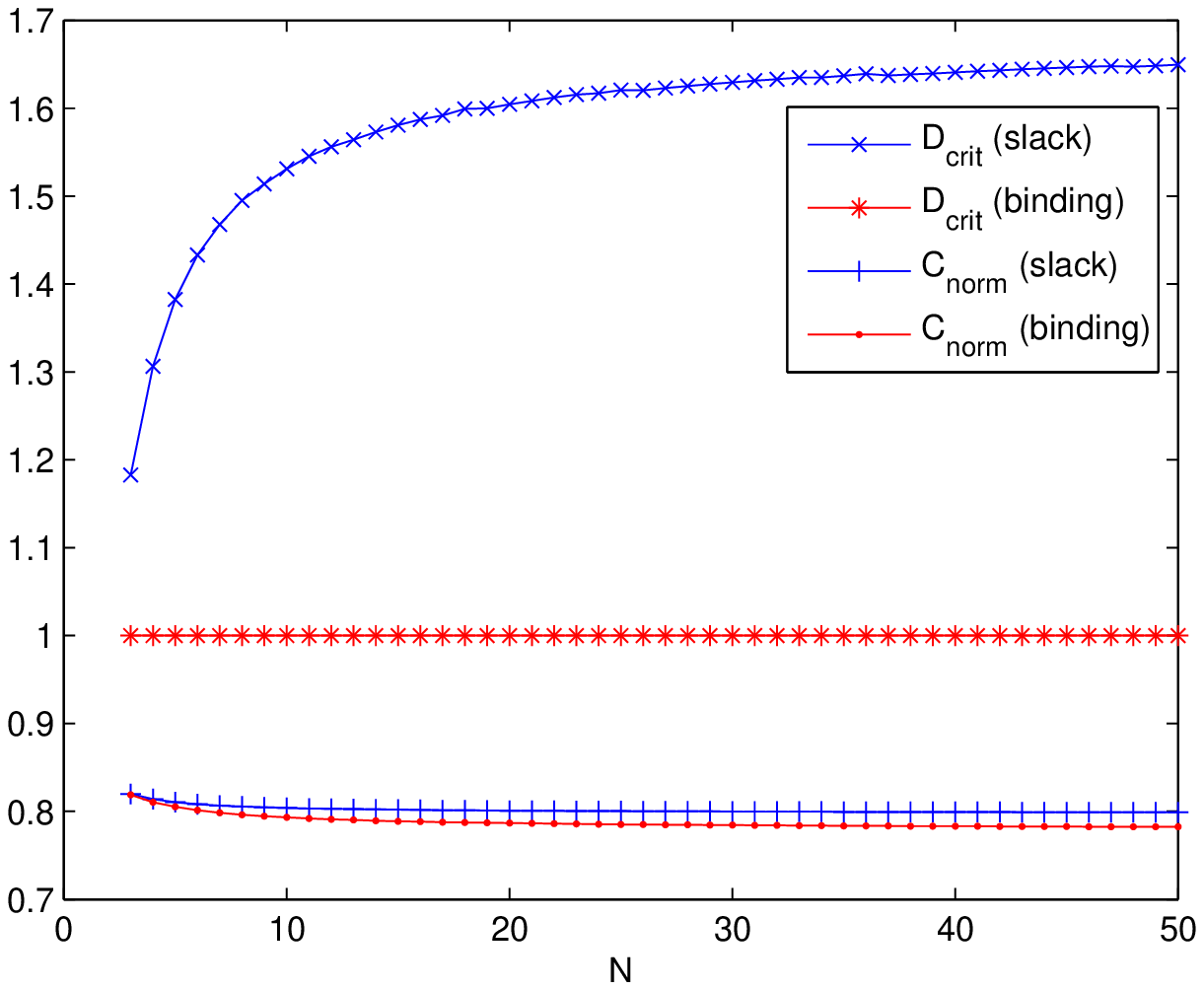}}%
\caption{Solution to the protocol design problem
for $N$ between $3$ and $50$ when $\theta = 0.1$:
\protect\subref{fig:nunc-a} optimal protocols, and
\protect\subref{fig:nunc-b} the values of $D_{crit}$ and $C_{norm}$ at the optimal protocols.}
\label{fig:nunc}%
\end{figure}

We examine how the optimal protocol changes as the number of
users varies between $3$ and $50$. We fix $\theta = 0.1$ as before. We first solve the
protocol design problem with a slack constraint, assuming that $\eta$ is sufficiently large.
Fig.~\ref{fig:nunc}\subref{fig:nunc-a} shows optimal
protocols $(q^*,r^*)$ when the delay constraint is not binding. As $N$ increases from $3$ to $50$, $q^*$ decreases
from $0.34$ to $0.02$ while $r^*$ decreases from $0.49$ to $0.48$.
Fig.~\ref{fig:nunc}\subref{fig:nunc-b} plots the values of $D_{crit}$ and $C_{norm}$
at $(q^*,r^*)$. As $N$ increases from $3$ to $50$,
$D_{crit}$ increases from $1.18$ to $1.65$ while
$C_{norm}$ decreases from $0.82$ to $0.80$.
The results show that when the delay constraint is slack, the delay in a critical phase
increases at a diminishing rate as the number of users increases, while the channel utilization in a normal
phase remains almost constant. Almost constant $C_{norm}$ implies that
the proposed optimal protocols are capable of resolving contention among
users efficiently in a normal phase even if there are many users sharing the channel.
The values of $D_{crit}$ at $(q^*,r^*)$
can be interpreted as the minimum values of $\eta$ that make the delay constraint slack.

Now we set $\eta = 1$ so that the delay constraint is binding for all $N$ between $3$ and $50$.
Fig.~\ref{fig:nunc}\subref{fig:nunc-a} shows optimal
protocols $(q^o,r^o)$ when the delay constraint is given by $D_{crit} \leq 1$. As $N$ increases from $3$ to $50$, $q^o$ decreases
from $0.34$ to $0.02$ while $r^o$ decreases from $0.40$ to $0.18$.
Imposing the constraint $D_{crit} \leq 1$ limits the values of $q$ and $r$,
but it impacts $r$ more than $q$, i.e., $r^o < r^*$ and $q^o \approx q^*$ for given $N$,
due to the shape of the contour curves of $C_{norm}$ as illustrated in Fig.~\ref{fig:contmap}.
Fig.~\ref{fig:nunc}\subref{fig:nunc-b} plots the values of $D_{crit}$ and $C_{norm}$
at $(q^o,r^o)$. As $N$ increases from $3$ to $50$,
$D_{crit}$ stays at $1$, confirming that the constraint $D_{crit} \leq 1$ is binding,
while $C_{norm}$ decreases from $0.82$ to $0.78$.
We can see that requiring $D_{crit} \leq 1$ decreases the maximum values of $C_{norm}$
only slightly because the delay constraint with $\eta = 1$ is mild so that
the optimal protocols remain interior. If we impose a sufficiently strong constraint,
i.e., choose a small $\eta$, then we have the optimal protocol at the corner, $q^o < q^*$ and $r^o = \epsilon$,
and $C_{norm}$ is reduced significantly, as suggested in Fig.~\ref{fig:gvary}.

\subsection{Varying the Fairness Level}

\begin{figure}%
\centering
\subfloat[][]{%
\label{fig:tunc-a}%
\includegraphics[width=0.5\textwidth]{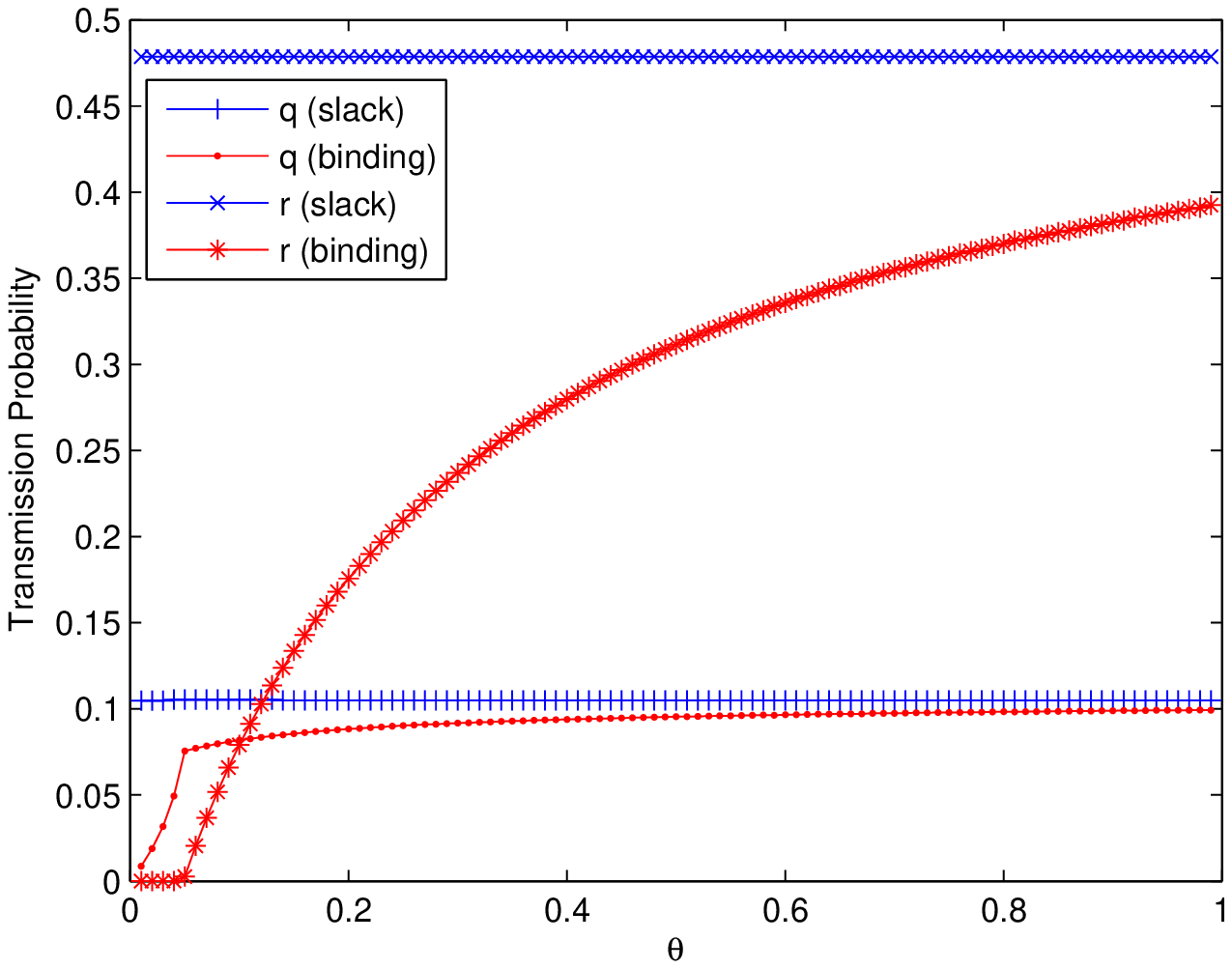}}%
\subfloat[][]{%
\label{fig:tunc-b}%
\includegraphics[width=0.5\textwidth]{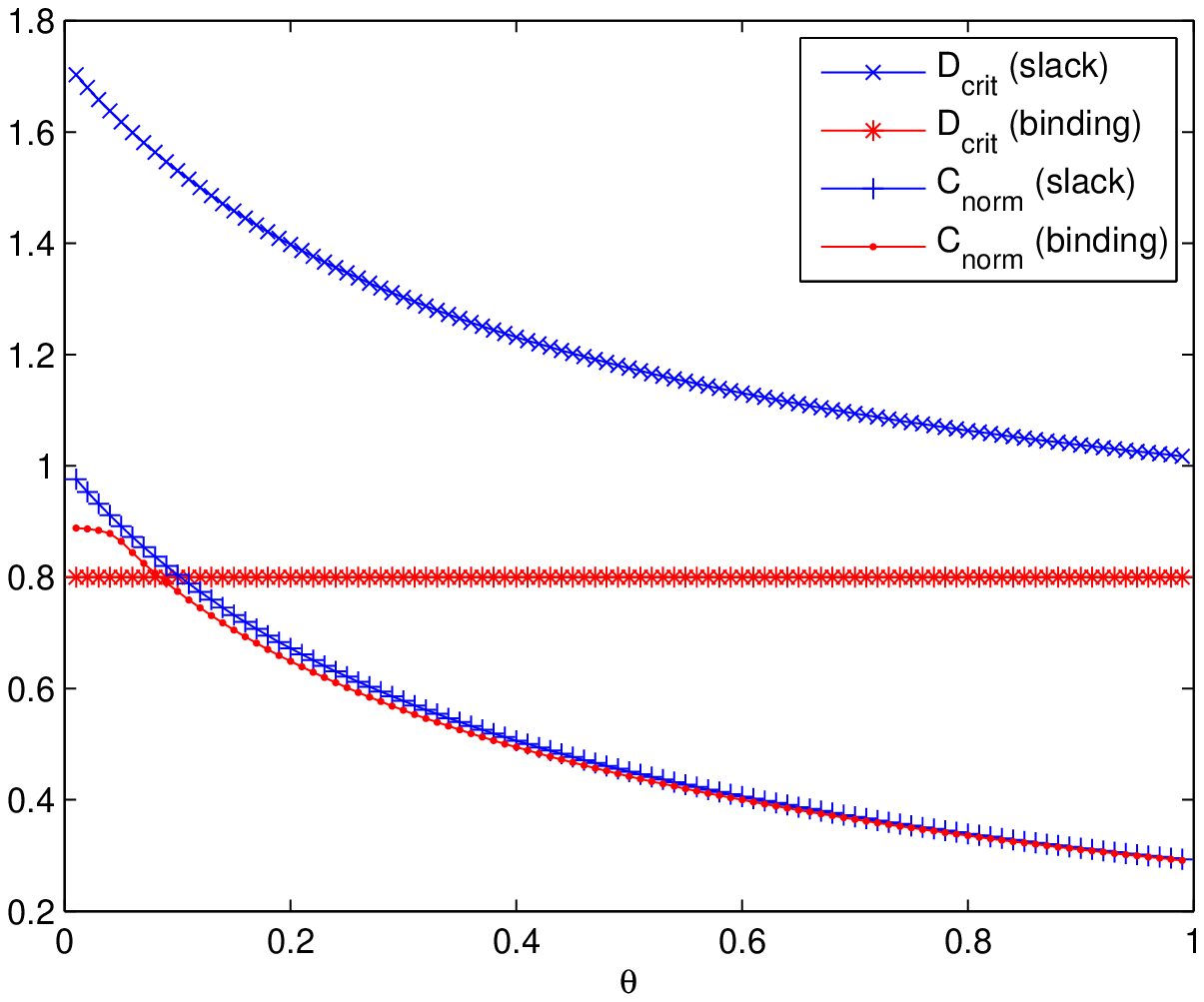}}%
\caption{Solution to the protocol design problem
for $\theta$ between $0.01$ and $0.99$ when $N = 10$:
\protect\subref{fig:tunc-a} optimal protocols, and
\protect\subref{fig:tunc-b} the values of $D_{crit}$ and $C_{norm}$ at the optimal protocols.}
\label{fig:tunc}%
\end{figure}

We investigate the impact of the fairness level
on optimal protocols and their performance. We first
consider sufficiently large $\eta$ so that the delay constraint is slack. Fig.~\ref{fig:tunc}\subref{fig:tunc-a} shows optimal
protocols $(q^*,r^*)$ when the constraint is slack. Since maximizing
$C_{norm}$ is equivalent to minimizing $T_c$, which is independent of
$\theta$, the optimal protocols do not depend on $\theta$ when the delay constraint is not binding.
Fig.~\ref{fig:tunc}\subref{fig:tunc-b} plots the values of $D_{crit}$ and $C_{norm}$
at $(q^*,r^*)$. From the expression in \eqref{eq:psuc}, we can see that $C_{norm}$ is decreasing in $\theta$.
$D_{crit}$ is also decreasing in $\theta$ because increasing $\theta$ induces
idle slots to occur more frequently, from which the average number of collisions experienced
by a critical user is small.

Since $D_{crit}$ at $(q^*,r^*)$ ranges between $1.02$
and $1.70$, we set $\eta = 0.8$ to analyze the protocol design
problem with a binding delay constraint. Fig.~\ref{fig:tunc}\subref{fig:tunc-a} shows optimal
protocols $(q^o,r^o)$ with $\eta = 0.8$ while Fig.~\ref{fig:tunc}\subref{fig:tunc-b} plots the values of $D_{crit}$ and
$C_{norm}$  at the optimal protocols. Note that the optimal protocols are at the corner
with $r^o = \epsilon$ for $\theta \leq 0.04$. Imposing the constraint $D_{crit} \leq 0.8$
limits the values of $q$ and $r$. The decrease in $q$ and $r$ is larger when $\theta$
is smaller because requiring $D_{crit} \leq 0.8$ imposes a stronger constraint
for smaller $\theta$, which can be seen by comparing the values of $D_{crit}$
with binding and slack delay constraints. However, the impact on $C_{norm}$ is marginal as long
as the optimal protocols are interior.

\subsection{Estimated Number of Users}

\begin{figure}%
\centering
\includegraphics[width=0.6\textwidth]{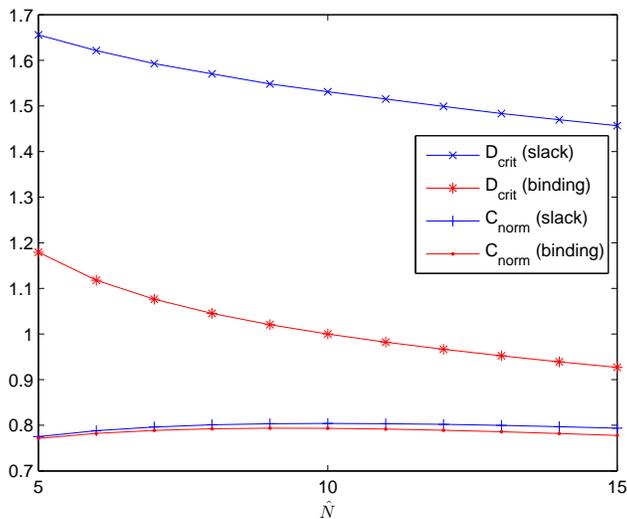}%
\caption{Values of $D_{crit}$ and $C_{norm}$ for $\hat{N}$ between 5 and 15
when $N = 10$ and $\theta = 0.1$.}
\label{fig:esti}%
\end{figure}

So far we have assumed that users know the exact number of users sharing
the channel. We relax this assumption
and consider a scenario where users follow optimal protocols computed
based on their (possibly incorrect) estimates of the number of users.
We investigate the consequence of using estimates
instead of the exact number of users when computing optimal protocols.
For simplicity, we assume that all users have the same estimate. We consider $N = 10$ and
the estimated number of users, denoted by $\hat{N}$, between $5$ and $15$.
In Fig.~\ref{fig:esti}, we plot the values of $D_{crit}$ and $C_{norm}$
when $N$ users follow the optimal protocol designed for $\hat{N}$ users. As before, we
consider the two cases of slack and binding delay constraints, with $\eta = 1$ for the binding
constraint. In both cases, optimal $q$ and $r$ decrease with the estimated number of users in order to
accommodate increased contention from more users, as shown in Fig.~\ref{fig:nunc}\subref{fig:nunc-a}.
Hence, $D_{crit}$ decreases with $\hat{N}$ since interference from normal users is reduced as $\hat{N}$ increases.
$C_{norm}$ is not affected much by $\hat{N}$, reaching a peak when $\hat{N} = N$.
This result suggests that the performance in a normal phase
is robust to errors in the estimation of the number of users.
Note that, in the case of the binding delay constraint, the constraint is violated when an
underestimation occurs, i.e., $\hat{N} < N$.
In order to avoid this, the protocol designer can choose an estimation procedure
that is biased toward overestimation. An estimation procedure can be designed
based on the approach of \cite{bianchi}, whose details are left for future work.

\section{Enhancement of Adaptive Protocols}

In this section, we discuss improvements on adaptive protocols
by utilizing longer memory. The main idea is the inference of the traffic types of other users
based on the patterns of observations. Some patterns of observations
reveal information about the types of other users, and users can adjust
their transmission parameters to these patterns. As users maintain longer
memory, there are more recognizable patterns, and exploiting them can achieve performance improvement.

\subsection{Reducing the Average Delay}

Returning to the discussion in Section IV.B, suppose that the transmission outcome
of the last slot of a normal phase is represented by $(1,W)$ so that there is a successful
user different from the critical user in the following critical phase.
When users follow a $\theta$-fair non-intrusive adaptive protocol, the successful user
transmits with probability $1-\theta$ in the first slot
of the following critical phase. If the successful user
transmits in the first slot, it collides with the critical user
and then transmits with probability $r$ in the next slot.
However, a collision cannot follow a success when all users have normal traffic, and thus
after a collision in the first slot of the critical phase, the successful user can infer the existence of
a critical user. If the protocol is modified so that it requires normal users to wait after
a pattern of $success$ followed by $failure$, then the average number
of collisions experienced by a critical user after a transmission outcome represented by $(1,W)$,
$d(1,W)$, is reduced from $(1-\theta)  [ (\mathbf{I} - \mathbf{Q}_{\textrm{\emph{crit}}})^{-1} \mathbf{e} ]_1$
to $1-\theta$. When $\theta$ is small, a success period lasts
long in a normal phase, leading to a large weight on $(1,W)$, $v(1,W)$. Thus, requiring normal users to wait after
$(success, failure)$ reduces the delay in a critical phase significantly.
For example, with $N = 10$, $\theta = 0.1$, and $(q,r) = (q^*,r^*) = (0.105, 0.479)$, we have
$d(1,W)$ reduced from $1.73$ to $0.9$, which decreases $D_{crit}$ from
$1.53$ to $0.93$.

\subsection{Bounding the Maximum Delay}

In the range of parameter values considered in Section V, the delay in a critical
phase is reasonably small, not exceeding 2 slots. However, the realized number of
collisions that a critical user experiences can be arbitrarily large
with positive probability. That is, the worst-case delay in a critical phase
is unbounded. We can bound the maximum delay by modifying the protocol so that it requires
normal users to wait after experiencing $B$ consecutive collisions. Since non-colliding
normal users wait after a collision, colliding normal users must have the same number
of consecutive collisions in any slot. Thus, normal users experiencing $B$ consecutive collisions
back off simultaneously, yielding a room for a critical user, if there is one.
Therefore, a critical user cannot experience more than $B$ collisions
in a critical phase. When $B$ is chosen moderately
large, $B$ consecutive collisions rarely occur in a normal phase, and thus
the proposed modification has a negligible impact on the channel utilization in a normal phase, $C_{norm}$.
We summarize below the enhanced adaptive protocols including the feature discussed in footnote 3.
\begin{enumerate}
\item If $y_i^{t-2} = success$, $y_i^{t-1} = failure$, and $z_i^t = normal$, then $p_i^t = 0$.

\item If $y_i^{t-B} = \cdots = y_i^{t-1} = failure$ and $z_i^t = normal$, then $p_i^t = 0$.

\item If $z_i^{t-1} = critical$ and $z_i^t = normal$, then $p_i^t = 0$.

\item Otherwise, $p_i^t = f(y_i^{t-1}, z_i^t)$, where $f$ is a $\theta$-fair non-intrusive adaptive protocol.
\end{enumerate}

\subsection{Two Users with Critical Traffic}

We now consider a scenario where the system can have two critical users at the same time.
We describe how the enhanced adaptive protocols can be extended to accommodate
two critical users.
Depending on the timing of the two arrivals of critical traffic, we analyze three cases where
two critical users coexist.

First, suppose that a second critical event occurs to user $j$ in slot $t$
while a critical user $i$ is having successful transmissions in a critical phase.
Since the traffic type of a user is its local information, user $j$
does not know whether or not there is another critical user in slot $t$.
We propose a protocol with which a user transmits when its traffic type
changes from normal to critical so that user $j$ transmits in slot $t$.
As in Section VI.A, the transmission by user $j$ informs user $i$ that there exists another
critical user in the system. If user $i$ had
normal traffic, it would respond by waiting in slot $t + 1$
according to the enhanced adaptive protocol so that user $j$ could capture the channel.
However, since user $i$ has critical traffic, we propose a protocol that makes user $i$ respond by
transmitting in slot $t + 1$ to inform user $j$ of its critical
traffic. Then after the implicit ``information exchange'' in slots $t$ and $t+1$,
both user $i$ and user $j$ know that there are two critical users. From slot $t + 2$ on, both users use the following decision rule
$g$ with initialization $idle$ to share the channel between them.
\begin{align*}
g(idle) = g(busy) = 1, \ g(success) = 0, \ g(failure) = 1/2.
\end{align*}
In slot $t + 2$, they collide, and after a collision, they transmit
with probability $1/2$.
Once one of the two users succeeds, they alternate between $(a_i, a_j) = (T, W)$
and $(a_i, a_j) = (W, T)$ until one of the users completes the transmission
of its critical traffic. After one of
the users completes the transmission of its critical traffic,
an idle slot occurs, and the situation becomes
the same as the one where a critical event arrives following an idle
slot. We can decrease the delay by requiring the user that
completed the transmission of its critical traffic earlier than the other
to wait in the slot following the idle slot.

Suppose now that two critical events occur simultaneously. Two critical users,
without knowing the existence of another critical user, transmit
with probability 1. After experiencing $B+1$ consecutive collisions, they
realize that another critical user exists because the maximum number of
consecutive collisions is bounded by $B$ when there is no or only one critical user
given the enhancement discussed in Section VI.B. Then after $B+1$ consecutive
collisions, the two critical users switch to the decision rule $g$ as above
in order to share the channel between them.

Lastly, suppose that a second critical event occurs to user $j$ in slot $t$ while a critical
user $i$ is experiencing collisions in a critical phase.
User $i$ realizes the existence of another critical user after $B+1$ consecutive collisions,
but at that point user $j$ has experienced less than $B+1$ consecutive collisions.
After experiencing $B+1$ consecutive collisions, user $i$ switches to the decision rule $g$ while user $j$ still transmits with
probability 1. From that point, there are only two possible transmission outcomes. Either user $j$ succeeds,
or users $i$ and $j$ collide.
If user $j$ experiences $B+1$ consecutive collisions before obtaining
a success, it switches to $g$ and the two users can share the channel from that point on. Suppose that
user $j$ obtains a success before experiencing $B+1$ consecutive collisions. Then it must be followed by
a collision because user $i$ uses $g$. Recognizing that the pattern $(success, failure)$
cannot occur when all other users have normal traffic, user $j$ also learns the existence
of another critical user and switches to $g$.

To summarize, when two critical users coexist, they can infer the
existence of another critical user within a finite number of slots,
recognizing the patterns that are not possible otherwise. Once the inference is made,
they switch to another mode of operation that enables them to share the channel equally.

\section{Simulation Results}

We have run simulations in order to confirm the results obtained in Sections IV and V
as well as the improvements from using an enhanced adaptive protocol introduced
in Section VI. We consider three values of $N$, $N = 3, 10, 50$, and three values of
$\theta$, $\theta = 0.1, 0.2, 0.5$. For each considered pair of $N$ and $\theta$, we have simulated
1,000 rounds of a normal phase for 100 slots followed by a critical phase (assuming only one critical user) while
choosing $(q,r)$ as the optimal protocol with a slack delay constraint, $(q^*,r^*)$.
Table~\ref{table:simul} summarizes the simulation results, showing the values of variables
$T_s$, $T_c$, $C_{norm}$, $D_{crit}$ averaged over 1,000 rounds as well as the maximum
value of $D_{crit}$ among the values in 1,000 rounds.
The results show that the simulation results match closely the results from analysis
in the case of adaptive protocols
and that enhanced adaptive protocols achieve a smaller delay in both average and
maximum senses without degrading the performance in a normal phase.
For the considered values of $N$ and $\theta$, $D_{crit}$ ranges from $0.68$ to
$1.02$ in the case of enhanced adaptive protocols and from $0.93$ to $1.66$
in the case of adaptive protocols. Hence, even without imposing a delay constraint,
we can achieve a reasonably small delay in a critical phase by using a
protocol proposed in this paper.
Also, $T_c$ ranges from $2.19$ to $2.55$, which shows that contention among
normal users is resolved effectively by a proposed protocol.

\begin{table}
\caption{Summary of simulation results (AP: adaptive protocols, EAP: enhanced adaptive protocols with $B = 5$)}
\centering
\begin{tabular}{|c|c||c|c c c c c|}
\hline
& & & $T_s$ & $T_c$ & $C_{norm}$ & $D_{crit}$ & $\max D_{crit}$ \\
\hline \hline
& & AP (analysis) & $10$ & $2.1959$ & $0.8199$ & $1.1786$ & \\
& $\theta = 0.1$ & AP (simulation) & $10.1727$ & $2.1865$ & $0.8146$ & $1.1820$ & $11$ \\
& & EAP (simulation) & $10.1761$ & $2.1879$ & $0.8145$ & $0.6820$ & $5$ \\
\cline{2-8}
$N=3$& & AP (analysis) & $5$ & $2.1959$ & $0.6948$ & $1.0899$ & \\
$(q^* = 0.3397,$ & $\theta = 0.2$ & AP (simulation) & $4.9567$ & $2.2122$ & $0.6872$ & $1.1340$ & $11$ \\
$r^* = 0.4896)$ & & EAP (simulation) & $4.9544$ & $2.2133$ & $0.6870$ & $0.7330$ & $5$ \\
\cline{2-8}
& & AP (analysis) & $2$ & $2.1959$ & $0.4767$ & $0.9352$ & \\
& $\theta = 0.5$ & AP (simulation) & $2.0105$ & $2.1853$ & $0.4788$ & $0.9340$ & $9$ \\
& & EAP (simulation) & $2.0106$ & $2.1876$ & $0.4786$ & $0.7660$ & $5$ \\
\hline
& & AP (analysis) & $10$ & $2.4374$ & $0.8040$ & $1.5297$ & \\
& $\theta = 0.1$ & AP (simulation) & $9.9352$ & $2.4316$ & $0.7953$ & $1.4480$ & $8$ \\
& & EAP (simulation) & $9.9365$ & $2.4358$ & $0.7952$ & $0.9180$ & $5$ \\
\cline{2-8}
$N=10$& & AP (analysis) & $5$ & $2.4374$ & $0.6723$ & $1.3978$ & \\
$(q^* = 0.1051,$ & $\theta = 0.2$ & AP (simulation) & $5.0662$ & $2.4562$ & $0.6696$ & $1.3910$ & $11$ \\
$r^* = 0.4786)$ & & EAP (simulation) & $5.0657$ & $2.4585$ & $0.6692$ & $0.9380$ & $5$ \\
\cline{2-8}
& & AP (analysis) & $2$ & $2.4374$ & $0.4507$ & $1.1759$ & \\
& $\theta = 0.5$ & AP (simulation) & $1.9926$ & $2.4350$ & $0.4501$ & $1.1180$ & $8$ \\
& & EAP (simulation) & $1.9928$ & $2.4367$ & $0.4499$ & $0.8930$ & $5$ \\
\hline
& & AP (analysis) & $10$ & $2.5138$ & $0.7991$ & $1.6468$ & \\
& $\theta = 0.1$ & AP (simulation) & $9.8745$ & $2.5389$ & $0.7872$ & $1.6570$ & $10$ \\
& & EAP (simulation) & $9.8807$ & $2.5452$ & $0.7870$ & $1.0230$ & $5$ \\
\cline{2-8}
$N=50$& & AP (analysis) & $5$ & $2.5138$ & $0.6654$ & $1.4995$ & \\
$(q^* = 0.0213,$ & $\theta = 0.2$ & AP (simulation) & $5.0067$ & $2.4998$ & $0.6615$ & $1.5090$ & $10$ \\
$r^* = 0.4754)$ & & EAP (simulation) & $5.0114$ & $2.5011$ & $0.6616$ & $1.0050$ & $5$ \\
\cline{2-8}
& & AP (analysis) & $2$ & $2.5138$ & $0.4431$ & $1.2546$ & \\
& $\theta = 0.5$ & AP (simulation) & $1.9945$ & $2.5180$ & $0.4418$ & $1.2470$ & $9$ \\
& & EAP (simulation) & $1.9950$ & $2.5193$ & $0.4417$ & $1.0060$ & $5$ \\
\hline
\end{tabular}
\label{table:simul}
\end{table}

\section{Conclusion}

We have explored the possibility of achieving coordination in a network with dynamically
changing user types by using adaptive MAC protocols with memory.
The general theme of this research agenda is to investigate the extent to which
the memory of local information can substitute explicit message passing
in achieving coordination.
In this paper, we are able to obtain satisfactory performance with protocols
utilizing short memory because we have focused on a relatively simple setting where
there are only two types and there can be at most one or two
critical users. In a more complex setting where there are more than two
types and more possible distributions of types, achieving coordination
by using only local information will become more difficult and, if possible, require longer
memory. We leave it as a future research topic to investigate the performance
of adaptive protocols with memory in a general setting of dynamically
changing user types.

\end{document}